\documentclass[journal]{IEEEtran}
\usepackage{graphicx}
\usepackage{xcolor}
\usepackage{amsmath}
\usepackage{authblk}
\usepackage{caption}
\usepackage{authblk}

\usepackage[sort,numbers]{natbib}

\begin{document}

\title{A New Experimental Method for Determining the Pinning Energy of an Abrikosov Vortex}

\author[1]{Elmeri O. Rivasto}
\affil[1]{CP3-origins, Department of Physics, Chemistry and Pharmacy, University of Southern Denmark, Campusvej 55, 5230 Odense, Denmark}

\maketitle

\begin{abstract}
We demonstrate a novel method of experimentally addressing the pinning energy of an Abrikosov vortex in high-temperature superconducting thin films. The method is based on our previously published theoretical framework considering the optimization of heterostructural bilayer films. To demonstrate our method, we have estimated the pinning energy of a typical BaZrO$_3$-nanocolumn within YBa$_2$Cu$_3$O$_{6+x}$ lattice using the experimental data from our previous study. Our calculations result in pinning energy of $0.04\,\mathrm{eV/nm}$ for a $c$-axis oriented vortex, which is in excellent agreement with a widely used theoretical formula for artificial columnar defects. However, the estimated energy is significantly higher than what has been previously reported for BaZrO$_3$-nanocolumns based on magnetic relaxation measurements. We argue that this discrepancy originates from the previously raised issues concerning the reliability of the magnetic relaxation measurements in determining the pinning energy. While the presented study only considers BaZrO$_3$-nanocolumns in YBa$_2$Cu$_3$O$_{6+x}$, the demonstrated method is equally applicable to any pinning site and superconducting material. The introduced method of measuring the pinning energy of Abrikosov vortices reduces the present ambiguity around the topic and enables more reliable modeling and design of YBCO based cryogenic memory cells and investigation for their use in quantum applications. 
\end{abstract}

\section{Introduction}
\noindent Abrikosov vortices are quantized units of magnetic flux ($\Phi_0 = 2.07\cdot 10^{-15}\,\mathrm{Wb}$) that emerge in the mixed state of type-II superconductors. While the interactions and dynamics of Abrikosov vortices have been widely studied over decades \cite{Blatter1994vortices}, their potential in various technological applications has only recently been acknowledged. The most notable application of Abrikosov vortices, whose proof of concept experiments have already been realized, is their use as classical bits in emerging cryogenic memory technologies \cite{Ortlepp2014access, Golod2015single, Alam2023cryogenic,Foltyn2024quantum, Karrer2024vortex}. Moreover, the possibility to manipulate and measure a single Abrikosov vortex has recently shown potential in quantum experiments and applications \cite{Foltyn2024quantum, Bak2024multi}. 

Knowing the pinning potential of the Abrikosov vortex is of utmost importance for designing and implementing both classical and quantum applications that take advantage of Abrikosov vortices. While vortex pinning has been widely studied using theoretical, computational and experimental methods \cite{Stoddart1993quantum, Shrivastava1993novel, Nelson1993boson, Blatter1994vortices, Sok1994thermal, Pan2001vortex, Palonen2012modeling, Assi2016disordered, Kwok2016vortices, Silva2017vortices, Paturi2018angle}, the absolute energy scale associated with vortices pinned in different defects remains unclear. Traditionally, the depth of the pinning potential (minimum energy, $u_0$) has been addressed via the superconducting condensation energy that equals to the energy associated with the Ginzburg-Landau thermodynamic critical field \cite{Blatter1994vortices, Zheng1994magnetic, Kwok2016vortices} 
\begin{equation}
    H_\mathrm{c} = \frac{\Phi_0}{2\sqrt{2} \lambda \xi},
\end{equation}
where $\lambda$ and $\xi$ are the (temperature dependent) London penetration depth and superconducting coherence length, respectively. In the case of a columnar defect of radius $R$ oriented parallel to the applied field (vortex core) and assuming that the $H_\mathrm{c}$ is confined within the $\xi$-radius core of the vortex and $\xi<R$, the pinning energy (per unit film thickness) can be calculated as \cite{Civale1991vortex, Civale1997vortex}
\begin{equation}
\label{theoretical_pinning_energy_Eq1}
    u_\mathrm{0} = -\frac{H_\mathrm{c}^2}{2\mu_0}\cdot \pi \xi^2,
\end{equation}
where $\mu_0=4\pi \cdot 10^{-7}\,$H/m is the magnetic permeability of the vacuum. In this work we will focus on studying the pinning energy associated with the widely studied BaZrO$_3$ (BZO) \cite{Goyal2005irradiation, Maiorov2009synergetic} columnar defects within YBa$_2$Cu$_3$O$_{6+x}$ (YBCO) lattice. The BZO nanocolumns grow along the YBCO $c$-axis and are associated with radius of $R=3\,\mathrm{nm}$ \cite{Wu2014effect, Shi2015influence}. In the low temperature limit the London penetration depth and superconducting coherence length along the YBCO $ab$-plane are generally considered to be $\lambda=150\,\mathrm{nm}$ and $\xi=1.5\,\mathrm{nm}$, respectively. With these values, Eq.~(\ref{theoretical_pinning_energy_Eq1}) results in pinning energy of $u_0 = 0.19\,\mathrm{eV/nm}$.

Another widely used formula for the pinning energy of large diameter columnar defects is given by \cite{Nelson1993boson, Blatter1994vortices, Klaassen2001vortex}
\begin{equation}
\label{theoretical_pinning_energy_Eq2}
    u_0 = -\frac{\Phi_0^2}{8\pi \mu_0 \lambda^2} \ln\left( 1 + \frac{R^2}{2\xi^2} \right).
\end{equation}
As for Eq.~(\ref{theoretical_pinning_energy_Eq1}), there is some ambiguity related to the value of $\mu_0$, which significantly affects the resulting energy. Using the value of vacuum permeability for the example system considered previously, Eq.~(\ref{theoretical_pinning_energy_Eq2}) results in $u_0=0.04\,\mathrm{eV/nm}$. This is significantly (almost 4 times) smaller energy than what was evaluated using Eq.~(\ref{theoretical_pinning_energy_Eq1}), illustrating how undetermined the absolute values of pinning energies are from a theoretical perspective. 

\begin{table*}[h!]
\centering
\begin{tabular}{ c|c|c|c|c|c|c } 
Reference & Sample & Film thickness & Field & Temperature & Reported energy & $u_0$ \\
         &              &  (nm)     &    (T)   &  (K) &   (K)  & ($\mu$eV/nm)\\
\hline
\hline
Miu \cite{Miu2012vortex, Miu2012non} & YBCO+BZO & 317 & 0.2 & 1 & 300  &  82 \\
                                     & -- & -- & 0.2 & 10 & 1000\  &  272  \\
                                     & -- & -- & 0.2 & 20 & 790  & 215   \\
                                     & -- & -- & 0.2 & 30 &  800 &   218 \\
                                     & -- & -- & 0.2 & 40 & 1000  &  272  \\
                                     & -- & -- & 0.2 & 50 &  1200 &  326  \\
                                     & -- & --& 3 & 1 & 150 &   43  \\
                                     & -- & -- & 3 & 10 & 450 &   122   \\
\hline
Miu \textit{et al.} \cite{Miu2010origin} & YBCO+BZO & 317 & 1  & 1  & 100 & 27 \\
                                         & --       & --  & -- & 5  &  300 & 82 \\
                                         & --       & --  & -- & 10 & 500 & 136 \\
                                         & --       & --  & -- & 20 & 700 & 190 \\
                                         & --       & --  & -- & 30 & 900 & 245 \\
                                         & --       & --  & -- & 40 & 1100 & 300 \\
\hline
Ijaduola \textit{et al.} \cite{Ijaduola2012critical} & NdBa$_2$Cu$_3$O$_{7-\delta}$+BZO & 700 & 0.5 & 0 & 70  &   9 \\
& -- & 2100 & 0.5 & 0 & 110\ &  5 \\
& -- & 700 & 1 & 0 & 200  &  25 \\
& -- & 2100 & 1 & 0 & 220  &  10  \\
\hline
Ivan \textit{et al.} \cite{Ivan2016vortex} & YBCO+4\%BZO+1\%Y$_2$O$_3$ & 400 & 0.2 & 20 & 750 &  162 \\
                                           & -- & -- & 0.2 & 30 & 840 & 181  \\
                                           & -- & -- & 0.2 & 40 &  1350 & 291  \\
                                           & -- & -- & 0.2 & 50 & 1450 &  312 \\

\end{tabular}
\caption{The experimentally determined pinning energies for the most relevant found studies regarding BZO-nanocolumns within high-temperature superconducting films. Most of the reported energies (in the units of K) have been visually estimated from the graphs presented in the associated publications.}
\label{pinning_energy_reference_table}
\end{table*}

Previous experimental measurements of the pinning energy have mostly relied on magnetic relaxation experiments. This means zero-field cooling the superconductor, ramping up the magnetic field and then measuring  $S=dM/d(\ln{t})$ \cite{Yeshurun1996magnetic}, where $M$ stands for magnetization. The unpinning (hopping) time of a vortex follows the Arrhenius relation $t = t_0 \cdot \exp(u /kT)$, where the pinning energy depends on the current density. In the simplest approximation, often referred to as the Anderson-Kim model \cite{Anderson1962theory, Anderson1964hard}, a linear dependence $u = u_0 \left( 
1 -J/J_c \right)$ is assumed. Solving the Arrhenius relation for $J$ using the linear approximation results in $J=J_c \left[ 1 - kT \ln{(t/t_0)}/u_0 \right]$. Since $M\propto J$ according to Bean's critical state model \cite{Bean1962magnetization, Poole2014superconductivity}, one can directly relate the experimentally measured magnetic relaxation rate $S$ with $J_\mathrm{c}$. In order to eliminate parameters from the equations that are difficult to address, one has to consider a normalized magnetic relaxation rate $S=\left(1/M \right) \cdot dM/d(\ln{t})$. Replacing $M$ by $J$ and inserting the above derived formula for $J$, one obtains
\begin{equation}
\label{magnetic_relazation_pinning_energy}
    u_0 = -\frac{kT}{S}.
\end{equation}
This is the simplest relation between experimentally measured $S$ and the underlying pinning potential. It is evident that the knowledge of $u(J)$ is essential for estimating the pinning potential using magnetic relaxation measurements. There exists more advanced models for the $u(J)$, in particular for high-temperature superconductors, resulting in more complicated relations between $u_0$ and $S$ \cite{Maley1990dependence, Fisher1989vortex, Feigel1989theory, Ijaduola2012critical}. 

Several previous studies have reported pinning energies for BZO-nanocolumns in high-temperature superconducting thin films calculated using the above described magnetic relaxation experiments under various field and temperature ranges \cite{Miu2012vortex, Miu2012non, Miu2010origin, Ijaduola2012critical, Ivan2016vortex}. In addition, the pinning energy has been recently measured by Ivan \textit{et al.} \cite{Ivan2022pinning} using the newly developed frequency-dependent AC susceptibility measurements. The results of these studies are listed in Table \ref{pinning_energy_reference_table}, where most of the estimated pinning energies at low temperatures can be observed to be 10--100 times smaller than what we obtained from Eq.~(\ref{theoretical_pinning_energy_Eq2}).

Despite its popularity, Yeshurun \textit{et al.} \cite{Yeshurun1996magnetic} have pointed out several concerns regarding the reliability of determining $u_0$ based on the magnetic relaxation measurements. Firstly, this requires one to assume a specific $u_0(J)$ relationship making it a highly model dependent approach. Related to this, the analysis assumes a periodic vortex lattice which can significantly deviate from the reality if the applied magnetic field in the relaxation experiments is not sufficient to establish fully penetrated flux distribution. This can make the chosen critical state model's assumed relationship between magnetization and critical current invalid. Secondly, the presence of magnetic field inhomogeneities in the widely used SQUID-based magnetometers cause the sample to experience significant hysteresis while being repeatedly scanned through the SQUID loop. This can result in an effect that resembles the AC demagnetization process, ultimately degrading the observed magnetization \cite{Yeshurun1996magnetic}. The evaluated pinning energies presented in Table~\ref{pinning_energy_reference_table} should thus be subject to critical scrutiny.

The presented introduction shows that there is significant discrepancy between the theoretically proposed and experimentally measured values of $u_0$. In order to resolve the present ambiguity, we will demonstrate the use of a novel experimental approach to determine the $u_0$, originally proposed in our previous works \cite{Rivasto2022optimization, Rivasto2023enhanced}. The method is based on growing a set of heterostructural bilayer films with fixed thicknesses, but alternating ratios ($f$) for the thicknesses of the individual layers. The bottom layer facilitates the doped YBCO where the pinning centers whose $u_0$ one wants to measure are present, while the top layer is pure YBCO that is ideally free of any pinning centers. The $u_0$ can then be estimated by measuring the zero-field $J_\mathrm{c}$s of the bilayers as a function of $f$ and determining the optimal $f$ resulting in highest $J_\mathrm{c}$. 

The manuscript is organized as follows: In section \ref{shape_of_pinning_potential_section} we address the shape of the pinning potential which we later need for evaluating the $u_0$. In section \ref{pinning_energy_scale_section} we give a detailed introduction to the method of calculating $u_0$ \cite{Rivasto2022optimization} and evaluate it for BZO-nanocolumns within YBCO at 10\,K temperature using our previously measured experimental data \cite{Rivasto2023enhanced}.

\section{Results and Discussion}
\subsection{Shape of the pinning potential}
\label{shape_of_pinning_potential_section}
\noindent In the following section the pinning potential will be evaluated based on the equality between pinning and Lorentz forces. To do this, one is required to know the shape of the pinning potential ($u$) as a function of distance between the center of the vortex and pinning center ($x_\mathrm{v}$), since the maximum pinning force is determined by the maximum slope of $u(x_\mathrm{v})$. In our previous work \cite{Aye2024enhanced}, we have used a generalized approach for calculating this function. Here, we will recap the basis of this approach and modify the associated calculations slightly in order to obtain more general results.

The spatial distribution of the superconducting order parameter ($\psi(x)$) in the vicinity of an Abrikosov vortex can be described by a variational solution of the Ginzburg-Landau equations \cite{Ginzburg2009theory} as $\psi_\mathrm{v}(x)=\psi_\mathrm{v,0}(x) \cdot \mathrm{e}^{i\theta}$, where \cite{Clem1975simple}
\begin{equation}
    \label{vortex_core_Eq}
    \psi_\mathrm{v,0}(x) =  \frac{x}{\left( x^2 + 2\cdot \xi(T)^2 \right)^{1/2}} ,
\end{equation}
and $\xi(T)= \xi_0 \cdot (1-T/T_\mathrm{c})^{-1/2}$ \cite{Blatter1994vortices}. To calculate the associated interaction energy, we use the generalized pinning energy formula (units: J/m) \cite{Aye2024enhanced}
\begin{equation}
\label{general_pinning_potential_Eq}
    u(x_\mathrm{v}) \propto \int_{-\infty}^{+\infty} \chi(x)\cdot \left|\psi_\mathrm{v}(x-x_\mathrm{v}) \right|^2 dx,
\end{equation}
where the dimensionless \textit{form function} $\chi(x)$ relates to the spatial variation of $\psi$ in the vicinity of a columnar defect, that is explicitly $\chi(x)=1-|\psi(x)|^2$. According to the two-fluid model of superconductivity, the local degradation of $\psi$ gives rise to a non-zero magnetic permeability ($\mu$). It then follows that $\chi(x)\propto \mu(x)$, which is an intuitively more convenient way to think about the physical interpretation of $\chi(x)$. 

While Eq.~(\ref{general_pinning_potential_Eq}) is equivalent to the previously used formulas for the pinning potential \cite{Blatter1994vortices, Klaassen2001vortex, Muller2001marginal, Pashitskii2002pinning}, the introduction of $\chi(x)$ generalizes the formula to consider, in particular, the effects related to the temperature dependent regeneration of $|\psi(x)|^2$ from 0 to $\psi_0$ (ideally $\psi_0=1$) from the pinning site--superconductor interface. The length scale over which this happens is determined by $\xi$ via the equation $\psi(x') = \psi_0 \left[ 1 - \tanh\left( x'/( \sqrt{2}\xi(T) ) \right)  \right]$ \cite{Poole2014superconductivity}, where $x'$ is the distance from the local minimum of $\psi(x)$. The explicit formula for the form function describing a cylindrical cavity of radius $R$ centered at $x=0$ consequently takes the form
\begin{equation}
\label{Chi_eq}
\chi(x)=
\begin{cases}
1, & \text{if $|x|\leq R$}\\
1 - \tanh^2\left( \frac{|x-R|}{ \sqrt{2}\cdot \xi(T)} \right),& \text{otherwise}.
\end{cases}
\end{equation}
The increasing $\xi(T)$ broadens the effective size of the pinning center with temperature. Thus it becomes evident that the proposed generalization of the pinning potential is required to accurately address the shape of the $u(x_\mathrm{v})$ interaction, especially when temperature dependent properties are of concern.

\begin{figure}[t!]
\begin{center}
\includegraphics[width=8cm]{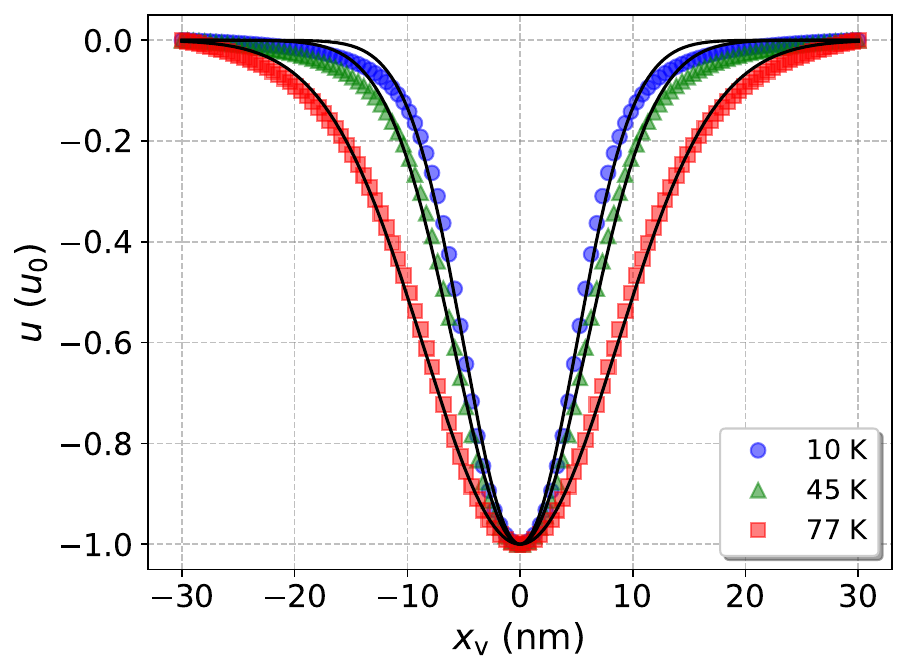}
\caption{Numerically calculated pinning potentials at 10\,K, 45\,K and 77\,K temperatures. The solid lines correspond to resulting fits of a Gaussian function $G(x_\mathrm{v})= \mathrm{exp}(-\gamma x_\mathrm{v}^2)-1$ via a single fitting parameter $\gamma$.}
\label{pinning_potential_Fig}
\end{center}
\end{figure}

We have numerically calculated $u(x_\mathrm{v})$ curves resulting from a BZO-nanocolumn ($R=3\,\mathrm{nm}$) located at $x=0$ at 10\,K, 45\,K and 77\,K temperatures. The results, together with fits of a Gaussian function of the form $G(x_\mathrm{v})= \mathrm{exp}(-\gamma x^2)-1$, are presented in Fig.~\ref{pinning_potential_Fig}. The spread of $u(x_\mathrm{v})$ as a function of temperature is clearly present. Nonetheless, the Gaussian function aligns well with the numerical data in the vicinity of the center of the nanocolumn at all temperatures. However, significant deviations between the fits and the calculated data can be seen for $x \gg R$ as the fits level of to zero much faster than the calculated points. The region associated with strongest pinning force (highest $|\partial_x u(x_\mathrm{v})|$) is, however, clearly in the region where the approximation of Eq.~(\ref{general_pinning_potential_Eq}) by a Gaussian is valid. As we will be restricting our analysis to the low temperature limit, we will only report the optimal fitting parameter $\gamma = 20 \cdot 10^{-3} \pm 2 \cdot 10^{-4} \,\mathrm{nm}^{-2}$ obtained at 10\,K. This value will be used in the following section to determine the energy (per unit film thickness) of an Abrikosov vortex pinned within a BZO-nanocolumn within the YBCO matrix at 10\,K temperature. 

\subsection{Determining the pinning energy}
\label{pinning_energy_scale_section}
\begin{figure}[t!]
\begin{center}
\includegraphics[width=7cm]{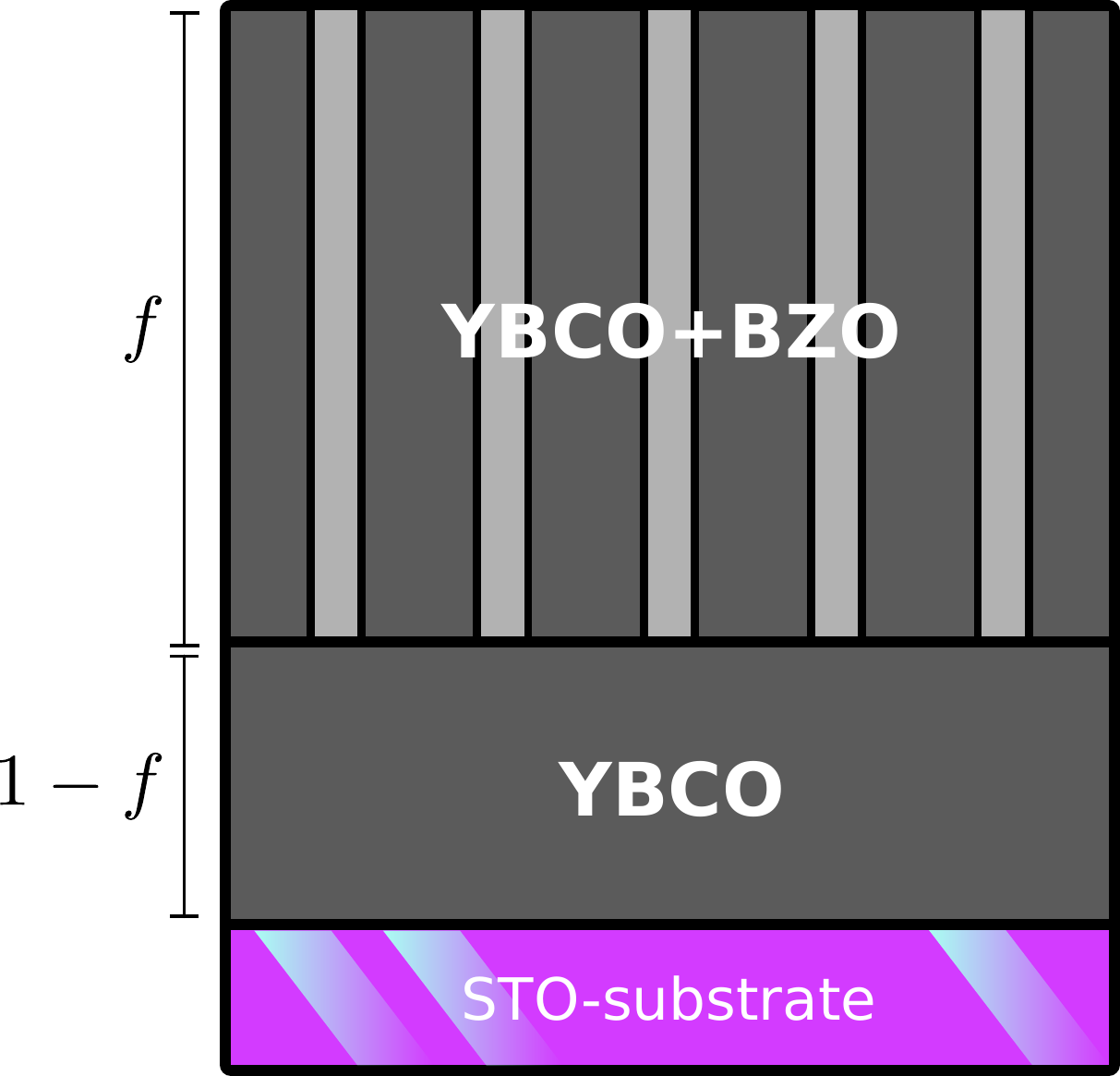}
\caption{A schematic illustration of the considered bilayer structure, the experimental realization of which has been presented in \cite{Rivasto2023enhanced}. }
\label{bilayer_schematics_Fig}
\end{center}
\end{figure}

\noindent With the knowledge of the shape of the pinning potential from the previous section, we can now proceed to evaluate the $u_0$ based on our previously published theoretical framework \cite{Rivasto2022optimization} and using the experimental data from our other closely related study \cite{Rivasto2023enhanced}. The main goal of these two studies was to find the optimal ratio for adjacent layers in a heterostructural bilayer film. The studied bilayers consisted of an intrinsic YBCO layer adjacent to a layer of BZO-doped YBCO grown on a single crystal SrTiO$_3$ (STO) substrate. The BZO forms nanocolumns that penetrate through the whole thickness of the doped layer as illustrated in Fig. \ref{bilayer_schematics_Fig}. From this onward, we will refer to the  relative thickness of the BZO-doped YBCO layer with an unitless parameter $f$. That is, for fully intrinsic and doped single layer films $f=0$ and $f=1$, respectively. 

To recap Ref.~\cite{Rivasto2022optimization}, the underlying idea leading to the estimation of $u_0$ is the following; The overall critical current density ($J_\mathrm{c}$) is limited either by the \textit{zero-field} $J_\mathrm{c}$ ($J_\mathrm{c,0}$) or the \textit{vortex dynamics limited} $J_\mathrm{c}$ ($J_\mathrm{c,v}$). The $J_\mathrm{c,0}$ is determined by the intrinsic properties of the superconductor related to the nanostructure of the associated material. Improving the homogeneity of the superconducting lattice increases $J_\mathrm{c,0}$ via maximizing the effective $|\psi|^2$ determining the number of Cooper pairs flowing through the sample. In consequence, the inclusion of any impurities, such as artificial pinning centers, inevitably decreases the $J_\mathrm{c,0}$. 

The simplest way to model the $J_\mathrm{c,0}(f)$ would be to assume that the intrinsic $J_\mathrm{c,0}$ of YBCO is not affected by the presence of any artificial pinning centers. Under this assumption the degradation of $J_\mathrm{c,0}$ with increasing $f$ manifests solely due to reduced superconducting cross-sectional area, resulting in linearly decreasing $J_\mathrm{c,0}(f)$. However, we have previously concluded that the substrate induced strain governed growth dynamics of YBCO on single crystalline SrTiO$_3$ results in non-linear decrease of $J_\mathrm{c,0}$ as a function of $f$. Thus, we have chosen to rely on purely experimental data originally presented in our previous study \cite{Rivasto2023enhanced} when considering the function $J_\mathrm{c,0}(f)$.

The $J_\mathrm{c,v}$ is simply determined as the (average) maximum current for which the vortices remain pinned within the superconductor. This is when the Lorentz force $f_\mathrm{L}=\Phi_0\cdot J$ resulting from the applied current $J$ equals to the maximum pinning force ($f_\mathrm{p,max}$) of the vortex (units N/m). It is evident that, contrary to the previously discussed $J_\mathrm{c,0}$, the $J_\mathrm{c,v}$ generally increases with the volumetric concentration of impurity material within the superconducting lattice due to enhanced vortex pinning. 

It is important to note that ideally $J_\mathrm{c,v}(f=0)=0$, since in a perfectly homogeneous superconducting lattice even the slightest Lorentz force resulting from the applied current will drive the vortices into motion causing power dissipation. The thermally activated vortex depinning rate can be approximated using Boltzmann statistics (Arrhenius expression \cite{Palstra1988thermally}) as $r_\mathrm{dp} \propto \exp\left( u_0(f)/kT \right) \approx u_0(f)/kT$, where $k$ is the Boltzmann constant and $T$ is the temperature. Since $J_\mathrm{c,v} \propto r_\mathrm{dp}$ and $u_0(f)\propto f$, the $J_\mathrm{c,v}$ can be concluded to increase linearly as a function of $f$, at least up to a first order approximation. This is further supported by our previous study where the vortex dynamics in the herein considered bilayer structures were simulated using molecular dynamics method \cite{Rivasto2022optimization}. Since $J_\mathrm{c,v}(f)$ is linear and we already concluded that $J_\mathrm{c,v}(f=0)=0$, one only requires to know a single value of $J_\mathrm{c,v}$ within the range $f\in (0,1]$ to determine an explicit formula for $J_\mathrm{c,v}(f)$. We will return to finding this formula shortly.
\begin{figure}[t!]
\begin{center}
\includegraphics[width=8cm]{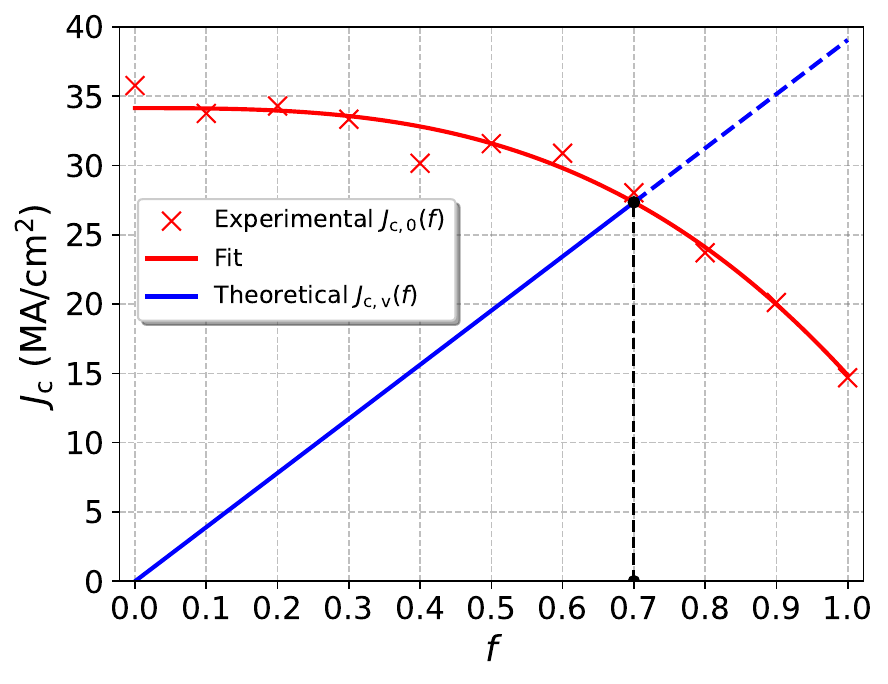}
\caption{The determination of $J_\mathrm{c,v}(f)$ based on the experimentally measured values of $J_\mathrm{c,0}(f)$ and associated optimal bilayer ratio of $f=0.7$ at 10\,K temperature (experimental data taken from Ref.~\cite{Rivasto2023enhanced}). The found optimum value for $f$ applies between 0.5--7\,T magnetic fields. }
\label{Jc0-Jcv_data_Fig}
\end{center}
\end{figure}

So far we have deduced that the $J_\mathrm{c,0}(f)$ and the $J_\mathrm{c,v}(f)$ are functions with decreasing and increasing slopes, respectively. Since the overall $J_\mathrm{c}$ must be limited by one of these, the optimal $f$ associated with the highest possible $J_\mathrm{c}$ for the associated bilayer structure is found at the point $f=f_\mathrm{opt}$ where the functions $J_\mathrm{c,0}(f)$ and $J_\mathrm{c,v}(f)$ intersect. Written explicitly;
\begin{equation}
\label{f_opt_Eq}
    J_\mathrm{c,0}(f_\mathrm{opt}) = J_\mathrm{c,v}(f_\mathrm{opt}),\texttt{ } \mathrm{where}\texttt{ } f_\mathrm{opt}\in [0,1].
\end{equation}
Since the $J_\mathrm{c,0}(f)$ can be found using direct experimental measurements, the only requirement for deriving the value for the $f_\mathrm{opt}$ is the knowledge of $J_\mathrm{c,v}(f)$. Vice versa, knowing $f_\mathrm{opt}$ enables one to estimate $J_\mathrm{c,v}(f)$ (under the assumption that $J_\mathrm{c,v}(f=0)=0$). The determination of $J_\mathrm{c,v}(f)$ is illustrated in Fig. \ref{Jc0-Jcv_data_Fig}, presenting a fit of a purely empirical function $J_\mathrm{c,0}(f)=J_\mathrm{c,0} - a\cdot f^n $ to the corresponding experimental data measured at 10\,K temperature \cite{Rivasto2023enhanced}. In Ref.~\cite{Rivasto2023enhanced} we concluded that the highest $J_\mathrm{c}$ of $27\cdot 10^{10}$\,A/m$^2$ was achieved with optimal bilayer ratio $f_\mathrm{opt}=0.7$. Given the measured values of $J_\mathrm{c,0}(f)$ \cite{Rivasto2023enhanced}, the equation for $J_\mathrm{c,v}(f)$ can now be deduced. This is illustrated in Fig.~\ref{Jc0-Jcv_data_Fig}, further showing how the value of $J_\mathrm{c,v}(f=1)\approx 39\cdot 10^{10}$\,A/m$^2$ was extrapolated. 

Next, we will deduce how to convert the estimated value of $J_\mathrm{c,v}(f=1)$ to $u_0$. As concluded in section \ref{shape_of_pinning_potential_section}, a Gaussian function is a very good approximation for the pinning potential of a BZO-nanocolumn in the region where the slope $|\partial_{x_\mathrm{v}} u(x_\mathrm{v})|$ reaches the maximum. This means that the maximum pinning force can be reliably evaluated as the maximum force resulting from the Gaussian pinning potential $u(x_\mathrm{v})= u_0\cdot \mathrm{exp}(-\gamma x_\mathrm{v}^2)$, where $\gamma = 20 \cdot 10^{-3} \pm 2 \cdot 10^{-4} \,\mathrm{nm}^{-2}$ based on the performed fit to numerically calculated set of  $u(x_\mathrm{v})$s at 10\,K. The slope of the Gaussian reaches maximum at $x_\mathrm{max}=\pm 1/\sqrt{2\gamma}$, resulting in the maximum pinning force $\left|\partial_x u(x) |_{x_\mathrm{max}}\right|$ given explicitly by $f_\mathrm{p,max} = u_0 \sqrt{2\gamma/\mathrm{e}}$ where $\mathrm{e}$ is Napier's constant. At the vortex dynamics limited critical current $f_\mathrm{p,max}=f_\mathrm{L}$ \cite{Matsuura1991flux}, from which one can solve $u_0$ resulting in 
\begin{equation}
\label{u0_Eq}
    u_0 = \sqrt{\frac{\mathrm{e}}{2\gamma}} \cdot \Phi_0 J_\mathrm{c,v}(f=1) \approx 0.04\,\mathrm{eV/nm}.
\end{equation}
The obtained value for the $u_0$ corresponds precisely to the value evaluated using the theoretical Eq.~(\ref{theoretical_pinning_energy_Eq2}) for a similar system. This suggests that the assumed values of $\lambda=150\,\mathrm{nm}$ and $\xi=1.5\,\mathrm{nm}$ describe the superconducting state of the considered YBCO+BZO film up to a high accuracy and validates the particular use of $\mu_0 = 4\pi\cdot 10^{-7}\,\mathrm{H/m}$ for magnetic permeability in Eq.~(\ref{theoretical_pinning_energy_Eq2}). 

Determining the $u_0$ based on the condensation energy (Eq.~(\ref{theoretical_pinning_energy_Eq1})) significantly overshoots the pinning energy. On the contrary, the previous estimations of $u_0$ based on the magnetic relaxation experiments (Table \ref{pinning_energy_reference_table}) result in approximately 10--100 times lower values of $u_0$ than what we obtained here. This will likely result from the issues associated with the magnetic relaxation experiments raised in the introduction. The herein evaluated value of $u_0$ can be considered reliable given its good agreement with Eq.~(\ref{theoretical_pinning_energy_Eq2}) that has been widely used for modeling vortex dynamics in high-temperature superconducting films \cite{Klaassen2001vortex, Paturi2018angle, Rivasto2020self}. 

\section{Conclusions}
\label{conclusions_section}
\noindent We have demonstrated a novel experimental approach for determining the pinning energy of an Abrikosov vortex in high-temperature superconducting thin films. Using the experimental data previously published in Ref. \cite{Rivasto2023enhanced}, we have estimated the energy of an Abrikosov vortex pinned within a BaZrO$_3$-nanocolumn at 10\,K temperature to be $0.04\,\mathrm{eV/nm}$. The obtained pinning energy is in good agreement with a theoretical formula (Eq.~(\ref{theoretical_pinning_energy_Eq2})) that has been widely used for modeling of vortex dynamics in high-temperature thin films with artificial columnar defects. However, previous experimental magnetic relaxation studies have estimated the associated pinning energy to be 10--100 times lower. This discrepancy most likely results from the several raised issues considering the reliability of the magnetic relaxation measurements pointed out in Ref.~\cite{Yeshurun1996magnetic}. 

While we have only considered the pinning energies associated with BZO-nanocolumns within YBCO, the demonstrated experiment equivalently applies for any superconducting material and artificial pinning center. The newly enabled ability to address pinning energies allows for more reliable theoretical modeling and design of cryogenic memory cells based on various superconducting materials, particularly high-temperature superconductors. Our results should also be of interest in the investigation of the possible quantum mechanical nature of Abrikosov vortices and their potential application in quantum information storage.

\section{Acknowledgments}
E.O.R. would like to thank Manuel Meyer for valuable feedback and suggestions during the preparation of this manuscript. E.O.R. acknowledges support from the European Research Council (ERC) under the European Union’s Horizon 2020 research and innovation program Grant agreement No. 948689 (AxionDM).

\bibliography{bibliography.bib}
\bibliographystyle{abbrv}

\end{document}